\documentclass[aps,prl,reprint,superscriptaddress]{revtex4-2}
\usepackage{amsfonts}
\usepackage{amssymb}
\usepackage{amsmath}
\usepackage{graphicx}
\usepackage{epsfig}
\usepackage{array}
\usepackage{braket}
\usepackage{multirow}
\usepackage[table,xcdraw]{xcolor}
\usepackage[colorlinks=true, citecolor=blue, urlcolor=blue, linkcolor=red]{hyperref}

\begin{document}

\title{Linear exciton Hall and Nernst effects in monolayer two-dimensional
semiconductors}
\author{Weilong Guo}
\author{Lianguo Li}
\author{Qingjun Tong}
\author{Ci Li}
\email{lici@hnu.edu.cn}
\affiliation{School of Physics and Electronics, Hunan University, Changsha 410082, China}

\begin{abstract}
This paper focuses on the study of linear exciton Hall and Nernst effects in
monolayer two-dimensional (2D) semiconductors, employing the semi-classical
transport theory. By deriving the exciton Berry curvature in momentum space
for a general inhomogeneous 2D system, we establish its dependence on the
Berry curvature and the effective mass of electron and hole. As illustrative
examples, the exciton Hall effect in monolayer transition metal
dichalcogenides (TMDs) and black phosphorus (BP) are calculated. For these
materials, we demonstrate that a linear Hall (Nernst) exciton current with
the non-zero Berry curvature is strictly forbidden by the symmetries. This
finding aligns with earlier experimental observations on the exciton Hall
effect in MoSe$_2$. In contrast, a strong anisotropy in BP leads to a net
linear Hall current of excitons, exhibiting a relatively large value and
resembling an anomalous Hall effect rather than a valley Hall effect. Our
work reveals that the specific symmetry of 2D materials can induce a
significant linear exciton Hall (Nernst) effect even without Berry
curvature, which is normally forbidden with non-zero Berry curvature in the
monolayer 2D material. This observation holds promise for future
optoelectronic applications and offers exciting possibilities for
experimental exploration.
\end{abstract}

\maketitle


\section{Introduction}

Excitons, composed of conduction electrons and valence holes with Coulomb
exchange interaction, represent fundamental elementary excitations with
significant potential for optoelectronic applications \cite{Yu,Xu,Zhang,Urb}%
. Currently, transition metal dichalcogenides (TMDs) serve as the primary
platform for exciton research in two-dimensional (2D) semiconductors due to
their unique optical selection rules. Specifically, the time-reversal and
broken inversion symmetries in monolayer TMDs result in a direct bandgap
that lies within the visible to near-infrared wavelength range. This
positions the conduction and valence band edges at the doubly degenerate
corners ($\pm K$ valleys) of the hexagonal first Brillouin zone (BZ),
enabling interband transitions in the $\pm K$ valleys to couple with
opposite circularly polarized light. Consequently, excitons dominate the
optical properties of TMDs \cite{Yu,Xu,Zhang,Urb}, allowing for manipulation
of the pseudospin associated with the valleys \cite%
{Yu,Wang,Hei,Hao,Shan,Yu1,Lou}. The reduced electron screening in 2D
materials leads to high exciton binding energies (on the order of hundreds
of $\mathrm{meV}$), ensuring their stability even at room temperature. The
small Bohr radius of excitons in TMDs ($1\sim 2\,\mathrm{nm}$) enhances
their coupling strength with light, making TMDs suitable for optical and
optoelectronic devices such as lasers, light-emitting diodes, and
optoelectronic detectors \cite{Urb,Shan}. Recent studies have further
revealed that monolayer TMDs on patterned substrates exhibit ground-state
excitonic properties \cite{Yang,Yang1}, attributed to their strong
sensitivity to the surrounding dielectric environment \cite{Raj,Shan1}.

On the other hand, the physics of excitons in anisotropic 2D semiconductors,
such as black phosphorus (BP), adds a new dimension to optical manipulation.
Monolayer BP features a direct bandgap (approximately $1\sim 2\,\mathrm{eV}$%
) centered around the $\Gamma $ point in the first BZ \cite{Neto,Gom,Ji},
giving rise to anisotropic excitonic states \cite{Neto1} that couple to
linearly polarized light through an optical dipole moment \cite%
{Gom,LYang,XWang}. These unique properties make BP highly promising for
optoelectronic applications, including field-effect transistors \cite{YBZ},
actively tunable spectra \cite{Kim}, and pseudospin-selective Floquet band
engineering \cite{Zhou}.

Previous studies have extended the understanding of the correlation between
Berry curvature in electronic systems and quantum transport phenomena \cite%
{Niu4,Niu5,Niu6,Niu7,Niu,Niu8,Yao1} into the realm of bosonic systems. In
this context, neutral excitons can be influenced by an external
electromagnetic field \cite{Ima,Gol,She}. Specifically, excitons exhibit a
Hall or Nernst effect induced by a non-zero excitonic Berry curvature when
there is an applied external field \cite{Niu1,Wu,Iwa,Lun,Sav,Gao}, i.e., an
in-plane external field within a 2D sample can induce a transverse exciton
current due to a finite excitonic Berry curvature. When the external
perturbation corresponds to a driving force or a gradient of chemical
potential, the resulting transverse response is referred to as the exciton
Hall effect. When the driving perturbation is a temperature gradient, the
analogous transverse response is the exciton Nernst effect. This phenomenon
is analogous to the anomalous Hall effect observed in electronic systems.
Recent investigations on excitons in monolayer and twisted multi-layer TMDs 
\cite{Yu,Yu1,Xu1,Li} have revealed that the effective exciton model, based
on valley excitons, exhibits higher symmetry than the spatial symmetry of
the underlying 2D material. Compared to earlier works that focused on the
exciton Hall effect in 2D semiconductor heterostructures \cite%
{Wu,Lun,Sav,Gao} or studies based on electron-hole (e-h) exchange
interactions with screening effects \cite{Wu}, a fundamental question
arises: Does a linear exciton Hall (Nernst) effect exist in monolayer 2D
semiconductors? If so, what are the necessary conditions according to
transport theory?

In this paper, we address these questions by expanding on the previously
developed excitonic transport theory \cite{Niu1}. Using monolayer TMDs and
BP as representative examples, we demonstrate that the anisotropy of 2D
materials and symmetries near the high symmetric points in the first BZ are
critical factors contributing to the linear exciton Hall (Nernst) effect.
For TMDs, symmetry considerations in both real and momentum space forbid a
linear exciton Hall (Nernst) response. This conclusion is consistent with
previous experimental observations of the exciton Hall effect in MoSe$_{2}$ 
\cite{Iwa}. In BP, the symmetry around the $\Gamma $ point in the first
BZ enforces a vanishing exciton Berry curvature, which again
rules out a Berry-curvature-driven linear exciton Hall (Nernst) effect.

In sharp contrast, the strong in-plane anisotropy of BP gives rise to a
highly anisotropic effective exciton Hamiltonian. As a result, a sizable
transverse exciton current can emerge \textit{without} Berry curvature, in a
manner reminiscent of an \textquotedblleft anomalous Hall
effect\textquotedblright\ \cite{Note1} rather than a conventional valley
Hall effect. Our work shows that suitable symmetry configurations in 2D
materials can yield a substantial net linear exciton Hall current even in
the absence of Berry curvature, thereby opening new avenues for experimental
investigation.

The paper is organized as follows: In Section II, we introduce and expand
the linear transport theory for excitons in 2D semiconductors. Section III
provides analytical and numerical results on the exciton Hall effect in
monolayer TMDs and BP for the non-zero Berry curvature. Similar calculations
for these materials without Berry curvature are discussed in Section IV.
Finally, we summarize our findings in Section V.

\begin{figure}[tbp]
\begin{center}
\includegraphics[width=0.48\textwidth]{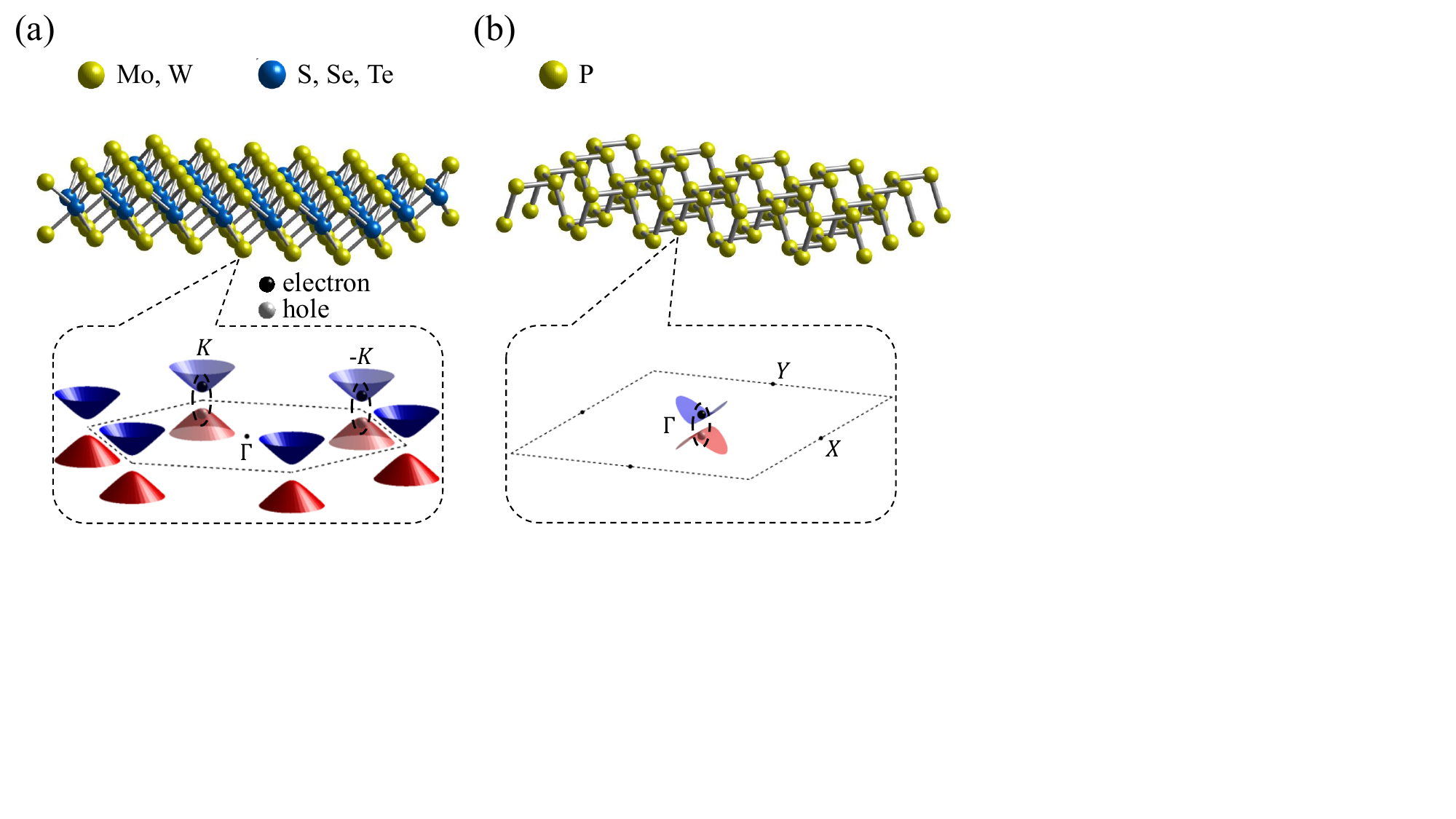}
\end{center}
\caption{(Color online) (a) Top panel: Schematic of the atomic structure of
a monolayer TMD. Bottom panel: Schematic of the band edge near these high
symmetric points ($\pm K$ valley) in the first BZ of a monolayer TMD. The
possible valley excitons are also shown within the black dashed circle. (b)
The similar plot to (a) but for monolayer BP.}
\label{fig1}
\end{figure}

\section{General description of the linear transport theory for excitons in
2D semiconductors}

\subsection{Symmetry analysis}

In a 2D electronic system, the conductivity $\boldsymbol{\sigma }$ can be
expressed as a $2\times 2$ tensor when only the linear response to an
external electric field $\boldsymbol{E}=\left( E_{x},E_{y}\right) $ is
considered \cite{Niu,Ong}. Ohm's law then takes the form $\boldsymbol{j}=%
\boldsymbol{\sigma E}$, i.e.,%
\begin{equation*}
\left( 
\begin{array}{c}
j_{x} \\ 
j_{y}%
\end{array}%
\right) =\left( 
\begin{array}{cc}
\sigma _{xx} & \sigma _{xy} \\ 
\sigma _{yx} & \sigma _{yy}%
\end{array}%
\right) \left( 
\begin{array}{c}
E_{x} \\ 
E_{y}%
\end{array}%
\right) ,
\end{equation*}%
where $\boldsymbol{j}=\left( j_{x},j_{y}\right) $ is the 2D electric current
density. A similar analysis can apply to exciton transport in 2D systems.
Let $\boldsymbol{j}$ now denote the exciton current density, and consider a
uniform in-plane driving field $\boldsymbol{\tilde{F}}=\left( \tilde{F}_{x},%
\tilde{F}_{y}\right) =\left( \tilde{F}\cos \theta ,\tilde{F}\sin \theta
\right) $, where $\theta $ is the angle between $\boldsymbol{\tilde{F}}$ and
the $x$-axis of the in-plane Cartesian coordinates. The driving field $%
\boldsymbol{\tilde{F}}$ may represent, for example, a mechanical force $%
\mathcal{F}$, a spacial gradient of chemical potential $\nabla \mu $, or a
spacial gradient of temperature $\nabla T$ in the material. In the
linear-response regime, one can write the exciton transport relation in
complete analogy with the electronic case: $\boldsymbol{j}=\boldsymbol{%
\sigma \tilde{F}}$.

It is often convenient to rotate the coordinate system such that the new
longitudinal axis $\parallel $ is aligned with the driving field $%
\boldsymbol{\tilde{F}}$. In the rotated frame $\left( \parallel ,\perp
\right) $,%
\begin{equation*}
\left( 
\begin{array}{c}
j_{\parallel } \\ 
j_{\perp }%
\end{array}%
\right) =\boldsymbol{\sigma }^{\prime }\left( 
\begin{array}{c}
\tilde{F} \\ 
0%
\end{array}%
\right) ,
\end{equation*}%
where $\boldsymbol{\sigma }^{\prime }$ is the conductivity tensor in the $%
\left( \parallel ,\perp \right) $ frame:%
\begin{equation*}
\boldsymbol{\sigma }^{\prime }=\left( 
\begin{array}{cc}
\sigma _{\parallel } & \sigma _{H}^{1} \\ 
\sigma _{H}^{2} & \sigma _{\perp }%
\end{array}%
\right) .
\end{equation*}%
By introducing the following symmetric (S) and antisymmetric (AS)
combinations:%
\begin{eqnarray}
\sigma _{\alpha \alpha }^{\mathrm{S}} &=&\sigma _{xx}+\sigma _{yy},\sigma
_{\alpha \alpha }^{\mathrm{AS}}=\sigma _{xx}-\sigma _{yy},  \label{S_AS} \\
\sigma _{\alpha \beta }^{\mathrm{S}} &=&\sigma _{xy}+\sigma _{yx},\sigma
_{\alpha \beta }^{\mathrm{AS}}=\sigma _{xy}-\sigma _{yx}.  \notag
\end{eqnarray}%
we obtain%
\begin{eqnarray}
\sigma _{\parallel } &=&\frac{\sigma _{\alpha \alpha }^{\mathrm{S}}+\sigma
_{\alpha \alpha }^{\mathrm{AS}}\cos 2\theta +\sigma _{\alpha \beta }^{%
\mathrm{S}}\sin 2\theta }{2},  \label{C_H} \\
\sigma _{H}^{1} &=&\frac{\sigma _{\alpha \beta }^{\mathrm{AS}}-\sigma
_{\alpha \alpha }^{\mathrm{AS}}\sin 2\theta +\sigma _{\alpha \beta }^{%
\mathrm{S}}\cos 2\theta }{2},  \notag \\
\sigma _{H}^{2} &=&\frac{-\sigma _{\alpha \beta }^{\mathrm{AS}}-\sigma
_{\alpha \alpha }^{\mathrm{AS}}\sin 2\theta +\sigma _{\alpha \beta }^{%
\mathrm{S}}\cos 2\theta }{2},  \notag \\
\sigma _{\perp } &=&\frac{\sigma _{\alpha \alpha }^{\mathrm{S}}-\sigma
_{\alpha \alpha }^{\mathrm{AS}}\cos 2\theta -\sigma _{\alpha \beta }^{%
\mathrm{S}}\sin 2\theta }{2}.  \notag
\end{eqnarray}%
Because 2D crystals can realize a variety of point-group symmetries, the
associated symmetry operations impose additional constraints on the form of
the conductivity tensor $\boldsymbol{\sigma }^{\prime }$, thereby
restricting which components (in particular, Hall-type responses) are
allowed or forbidden in a given material. We summarize these constraints on $%
\boldsymbol{\sigma }^{\prime }$, derived from time-reversal ($\mathcal{T}$)
symmetry \cite{Note2} together with ten crystallographic point groups in 2D, 
$C_{n=1,2,3,4,6}$ and $D_{n=1,2,3,4,6}$, in Table \ref{Table I} (see \cite{Supp}
for details of the derivation). In particular, the $D_{3h}$\ point-group
symmetry of monolayer TMDs forbids non-zero $\sigma _{H}^{1}$ and $\sigma
_{H}^{2}$, whereas the $D_{2h}$\ symmetry of monolayer BP allows all
components of $\boldsymbol{\sigma }^{\prime }$\ to be nonvanishing. In the
following, we illustrate and corroborate the symmetry-based results
summarized in Table \ref{Table I} from a complementary viewpoint, by
performing explicit momentum-space calculations for specific materials.

\begin{table*}[tbph]
\caption{Symmetry analysis of symmetric and antisymmetric combination $%
\protect\sigma _{\protect\alpha \protect\alpha \left( \protect\alpha \protect%
\beta \right) }^{\mathrm{S},\mathrm{AS}}$ [Eq. (\protect\ref{S_AS})] and
exciton conductivity tensor $\boldsymbol{\protect\sigma }^{\prime }$ [Eq. (%
\protect\ref{C_H})]. The analysis is based primarily on time-reversal ($%
\mathcal{T}$) symmetry \protect\cite{Note2} and ten crystallographic point
groups in 2D, i.e., $C_{n=1,2,3,4,6},D_{n=1,2,3,4,6}$ (see \protect\cite%
{Supp} for details). The table lists the additional constraints imposed by
these symmetries on the allowed components of $\protect\sigma _{\protect%
\alpha \protect\alpha \left( \protect\alpha \protect\beta \right) }^{\mathrm{%
S},\mathrm{AS}}$ and $\boldsymbol{\protect\sigma }^{\prime }$. The symbol
\textquotedblleft $\boldsymbol{-}$" denotes the absence of any
symmetry-imposed restriction on the corresponding component.}
\label{Table I}
\begin{center}
\setlength{\arrayrulewidth}{0.5mm} 
\renewcommand\tabcolsep{5.5pt} 
\begin{tabular}{ccccccccc}
\hline\hline
&  &  &  &  &  &  &  &  \\[-1ex] 
point-group symmetry with $\mathcal{T}$ & $\sigma^{\mathrm{S}}_{\alpha\alpha}
$ & $\sigma^{\mathrm{AS}}_{\alpha\alpha}$ & $\sigma^{\mathrm{S}%
}_{\alpha\beta}$ & $\sigma^{\mathrm{AS}}_{\alpha\beta}$ & $\sigma_{\parallel}
$ & $\sigma_{\perp}$ & $\sigma^1_{H}$ & $\sigma^2_{H}$ \\[0.5ex] \hline
&  &  &  &  &  &  &  &  \\[-1ex] 
$C_{2}$ & $\boldsymbol{-}$ & $\boldsymbol{-}$ & $\boldsymbol{-}$ & $0$ & $%
\boldsymbol{-}$ & $\boldsymbol{-}$ & $\boldsymbol{-}$ & $\boldsymbol{-}$ \\%
[0.5ex] 
$C_{n\geq 3}$ & $\boldsymbol{-}$ & $0$ & $0$ & $0$ & $\sigma_{xx}$ & $%
\sigma_{xx}$ & $0$ & $0$ \\[0.5ex] 
$D_{2}$ & $\boldsymbol{-}$ & $\boldsymbol{-}$ & $0$ & $0$ & $\boldsymbol{-}$
& $\boldsymbol{-}$ & $\boldsymbol{-}$ & $\boldsymbol{-}$ \\[0.5ex] 
$D_{n\geq 3}$ & $\boldsymbol{-}$ & $0$ & $0$ & $0$ & $\sigma_{xx}$ & $%
\sigma_{xx}$ & $0$ & $0$ \\[0.5ex] \hline
\end{tabular}%
\end{center}
\end{table*}

\subsection{General theory}

\vspace{-0.06cm}Based on the previous excitonic transport theory \cite{Niu1}%
, for a excitonic state around the high symmetric points (electronic band
edges) in the first BZ of the inhomogeneous 2D system, we can write this
state as%
\begin{equation}
\left\vert U_{l,\boldsymbol{k}}^{ex}\right\rangle =\sum_{\boldsymbol{q}%
}F_{l}\left( \boldsymbol{k},\boldsymbol{q}\right) e^{i\boldsymbol{q}\cdot 
\boldsymbol{r}}\left\vert u_{e,\boldsymbol{Q}_{e}}\right\rangle \left\vert
u_{h,\boldsymbol{Q}_{h}}\right\rangle ,
\end{equation}%
where $\boldsymbol{q}$ and $\boldsymbol{r}$ are, respectively, the wave
vector and coordinate for the relative motion. $u_{e}$ and $u_{h}$ are the
periodic part of the electron and hole Bloch function, and $\sum_{%
\boldsymbol{q}}F_{l}\left( \boldsymbol{k},\boldsymbol{q}\right) e^{i%
\boldsymbol{q}\cdot \boldsymbol{r}}$ gives the envelope function of the
relative motion which may depend on $\boldsymbol{k}$, the center-of-mass
(COM) momentum of excitons in general. $l$ is the quantum number for each
hydrogenlike orbit of the relative motion. $\boldsymbol{m}_{ex}=\left(
m_{ex,x},m_{ex,y}\right) =\left( m_{e,x}+m_{h,x},m_{e,y}+m_{h,y}\right) $ is
the exciton mass, i.e., the sum of effective electron and hole mass, with%
\begin{eqnarray}
\boldsymbol{Q}_{e} &=&\frac{m_{e,x}}{m_{ex,x}}k_{x}\hat{e}_{x}+\frac{m_{e,y}%
}{m_{ex,y}}k_{y}\hat{e}_{y}+\boldsymbol{q}, \\
\boldsymbol{Q}_{h} &=&\frac{m_{h,x}}{m_{ex,x}}k_{x}\hat{e}_{x}+\frac{m_{h,y}%
}{m_{ex,y}}k_{y}\hat{e}_{y}-\boldsymbol{q},  \notag
\end{eqnarray}%
the crystal momentum of electron and hole. In describing the exciton, it is
intuitive to equal the configuration of filled valence and empty conduction
bands to the vacuum, and use the hole operator to denote a vacancy in the
valence band as: $\left\vert u_{h,\boldsymbol{Q}_{h}}\right\rangle =\hat{h}_{%
\boldsymbol{Q}_{h}}^{\dagger }\left\vert 0_{h}\right\rangle =\hat{e}_{%
\boldsymbol{Q}_{h}}\left\vert 0_{h}\right\rangle \equiv \left\vert v_{%
\boldsymbol{-Q}_{h}}\right\rangle $, where $\left\vert 0_{h}\right\rangle $
means the vaccum of the hole, $\left\vert v\right\rangle $ represents the
electric state of valence band. The electron operator without the band index
then implicitly refers to that of the conduction electron, i.e., $\left\vert
u_{e,\boldsymbol{Q}_{e}}\right\rangle =\hat{e}_{\boldsymbol{Q}_{e}}^{\dagger
}\left\vert 0_{e}\right\rangle \equiv \left\vert c_{\boldsymbol{Q}%
_{e}}\right\rangle $.

So the Berry connection of this state in 2D limit can be expressed as $%
\mathcal{A}_{\boldsymbol{k}}^{ex}=i\left\langle U_{\boldsymbol{k}%
}^{ex}\right\vert \nabla _{\boldsymbol{k}}\left\vert U_{\boldsymbol{k}%
}^{ex}\right\rangle $, where we have ignored the index $l$ since only $l=0$
situation need to be considered in most of cases of 2D semiconductors \cite%
{Yu1,Lou}. Specifically,%
\begin{eqnarray}
\mathcal{A}_{k_{\alpha }}^{ex} &=&i\left\langle U_{\boldsymbol{k}%
}^{ex}\right\vert \partial _{k_{\alpha }}\left\vert U_{\boldsymbol{k}%
}^{ex}\right\rangle \\
&=&i\sum_{\boldsymbol{q}}[F^{\ast }\left( \boldsymbol{k},\boldsymbol{q}%
\right) \partial _{k_{\alpha }}F\left( \boldsymbol{k},\boldsymbol{q}\right) 
\notag \\
&&+\left\vert F\right\vert ^{2}\left( \frac{m_{e,\alpha }}{m_{ex,\alpha }}%
\left\langle u_{e}\right\vert \partial _{k_{\alpha }}\left\vert
u_{e}\right\rangle +\frac{m_{h,\alpha }}{m_{ex,\alpha }}\left\langle
u_{h}\right\vert \partial _{k_{\alpha }}\left\vert u_{h}\right\rangle
\right) ]  \notag \\
&\approx &i\sum_{\boldsymbol{q}}[F^{\ast }\left( \boldsymbol{q}\right)
\partial _{k_{\alpha }}F\left( \boldsymbol{q}\right)  \notag \\
&&+\left\vert F\right\vert ^{2}\left( \frac{m_{e,\alpha }}{m_{ex,\alpha }}%
\left\langle c\right\vert \partial _{Q_{e,\alpha }}\left\vert c\right\rangle
-\frac{m_{h,\alpha }}{m_{ex,\alpha }}\left\langle v\right\vert \partial
_{Q_{h,\alpha }}\left\vert v\right\rangle \right) ]  \notag \\
&=&i\sum_{\boldsymbol{q}}\left\vert F\right\vert ^{2}\left( \frac{%
m_{e,\alpha }}{m_{ex,\alpha }}\left\langle c\right\vert \partial
_{Q_{e,\alpha }}\left\vert c\right\rangle -\frac{m_{h,\alpha }}{m_{ex,\alpha
}}\left\langle v\right\vert \partial _{Q_{h,\alpha }}\left\vert
v\right\rangle \right)  \notag
\end{eqnarray}%
where $F\left( \boldsymbol{k},\boldsymbol{q}\right) \approx F\left( 
\boldsymbol{q}\right) $ is a good approximation for intralayer excitons in
2D materials \cite{Yu1,Lou}, leading to the Berry curvature as%
\begin{eqnarray}
\Omega ^{ex} &\equiv &\Omega ^{ex}\left( \boldsymbol{k}\right) =\nabla _{%
\boldsymbol{k}}\times \mathcal{A}_{\boldsymbol{k}}^{ex}=\partial _{k_{x}}%
\mathcal{A}_{k_{y}}^{ex}-\partial _{k_{y}}\mathcal{A}_{k_{x}}^{ex}  \notag \\
&\approx &\sum_{\boldsymbol{q}}\left\vert F\right\vert ^{2}\left[ \frac{%
m_{e,x}m_{e,y}}{m_{ex,x}m_{ex,y}}\Omega _{e}+\frac{m_{h,x}m_{h,y}}{%
m_{ex,x}m_{ex,y}}\Omega _{h}\right]  \notag \\
&\approx &\left[ \frac{m_{e,x}m_{e,y}}{m_{ex,x}m_{ex,y}}\Omega _{c}-\frac{%
m_{h,x}m_{h,y}}{m_{ex,x}m_{ex,y}}\Omega _{v}\right] \sum_{\boldsymbol{q}%
}\left\vert F\right\vert ^{2}  \notag \\
&=&\frac{\Omega _{c}\left( m_{e,x}m_{e,y}+m_{h,x}m_{h,y}\right) }{%
m_{ex,x}m_{ex,y}}\sum_{\boldsymbol{q}}\left\vert F\right\vert ^{2}
\end{eqnarray}%
Here $\Omega _{c}\left( \boldsymbol{Q}\right) =\Omega _{c}\left( \boldsymbol{%
k},\boldsymbol{q}\right) =\nabla _{\boldsymbol{Q}}\times i\left\langle
c\right\vert \partial _{\boldsymbol{Q}}\left\vert c\right\rangle =-\Omega
_{v}\left( \boldsymbol{Q}\right) $ is the Berry curvature of the conduction
(valence) band for the crystal momentum $\boldsymbol{Q}$ \cite{Note3}, which
can be calculated by the effective electric band-edge model of 2D materials.
Meanwhile, $\boldsymbol{Q}$ can be directly replaced by COM momentum $%
\boldsymbol{k}$ of excitons since $\Omega _{c\left( v\right) }\left( 
\boldsymbol{k},\boldsymbol{q}\right) \approx \Omega _{c\left( v\right)
}\left( \boldsymbol{k}\right) $, which comes from that the relative wave
vector $\boldsymbol{q}$ is always small comparing with $\boldsymbol{k}$ \cite%
{Yu1,Lou}.

According to previous theoretical works \cite{Niu1,Di,Xiao}, $\Omega
^{ex}\equiv \Omega _{n,\boldsymbol{k}}$ can be one of the source for the
current density of excitons in the intrinsic response%
\begin{eqnarray}
\boldsymbol{j} &=&\boldsymbol{j}_{0}+\boldsymbol{j}_{1}, \\
\boldsymbol{j}_{0} &=&\sum_{n}\int \frac{d^{2}\boldsymbol{k}}{\left( 2\pi
\right) ^{2}}f_{n}^{0}\left( \boldsymbol{k}\right) \boldsymbol{v}_{n}\left( 
\boldsymbol{k}\right) +\nabla \times \boldsymbol{M}\left( r\right) ,  \notag
\\
\boldsymbol{j}_{1} &=&\sum_{n}\int \frac{d^{2}\boldsymbol{k}}{\left( 2\pi
\right) ^{2}}f_{n}^{1}\left( \boldsymbol{k}\right) \boldsymbol{v}_{n}\left( 
\boldsymbol{k}\right) ,  \notag
\end{eqnarray}%
where $n$ is the band index for different excitonic bands,%
\begin{equation*}
\boldsymbol{v}_{n}\left( \boldsymbol{k}\right) =\frac{\partial \varepsilon
_{n}\left( \boldsymbol{k}\right) }{\hbar \partial \boldsymbol{k}}-\frac{%
\boldsymbol{\mathcal{F}}}{\hbar }\times \Omega _{n,\boldsymbol{k}}\hat{z}=%
\boldsymbol{v}_{b}+\boldsymbol{v}_{a},
\end{equation*}%
is the velocity of an exciton. $\boldsymbol{v}_{b}\equiv \frac{\partial
\varepsilon _{n}\left( \boldsymbol{k}\right) }{\hbar \partial \boldsymbol{k}}
$ means the band velocity, which is similar with the one for electronic
systems. $\boldsymbol{\mathcal{F}}=\hbar \frac{d\boldsymbol{k}}{dt}$
represents the in-plane driving force. We denote $\boldsymbol{v}_{a}$ as the
anomalous velocity arising from the non-zero excitonic Berry curvature. This
term plays a role analogous to an effective magnetic field in the
conventional Hall effect, generating a transverse exciton current
perpendicular to the applied force $\boldsymbol{\mathcal{F}}$.%
\begin{equation*}
\boldsymbol{M}\left( r\right) =\frac{k_{B}T}{\hbar }\sum_{n}\int \frac{d^{2}%
\boldsymbol{k}}{\left( 2\pi \right) ^{2}}\Omega _{n,\boldsymbol{k}}\hat{z}%
\log \left[ 1-e^{-\left( \varepsilon _{n}-\mu \right) /k_{B}T}\right] ,
\end{equation*}%
is analogous to the equilibrium magnetization density in electronic systems 
\cite{Niu1,Di}. Here $f_{n}^{0}\left( \boldsymbol{k}\right) =1/\left[
e^{\left( \varepsilon _{n}-\mu \right) /k_{B}T}-1\right] $ is the
equilibrium Bose-Einstein distribution function and $f_{n}^{1}\left( 
\boldsymbol{k}\right) =-\tau df_{n}^{0}\left( \boldsymbol{k}\right) /dt$.
Supposing there are $\boldsymbol{\mathcal{F}}=\mathcal{F}\left( \cos \theta
,\sin \theta \right) =\left( \mathcal{F}_{x},\mathcal{F}_{y}\right) $, $%
\nabla \mu $, and $\nabla T$ in the material (here we assume that $%
\boldsymbol{\mathcal{F}}$, $\nabla \mu $ and $\nabla T$ in the same
direction for simplicity), the previous derivation [Eq. (\ref{C_H})] shows
that the linear Hall (Nernst) current can be expressed as (see details of
derivation in \cite{Supp})%
\begin{eqnarray}
j_{\perp ,H} &=&\left[ \sigma _{yx}^{0}+\frac{\sin 2\theta }{2}\left( \sigma
_{xx}^{1}-\sigma _{yy}^{1}\right) \right] \mathcal{F}_{\parallel }  \notag \\
&&+\left[ \sigma _{yx}^{0}-\frac{\sin 2\theta }{2}\left( \sigma
_{xx}^{1}-\sigma _{yy}^{1}\right) \right] \partial _{\parallel }\mu  \notag
\\
&&+\left[ \alpha _{yx}^{0}+\frac{\sin 2\theta }{2}\left( \alpha
_{xx}^{1}-\alpha _{yy}^{1}\right) \right] \left( -k_{B}\partial _{\parallel
}T\right) ,
\end{eqnarray}%
with%
\begin{eqnarray}
\sigma _{xy}^{0} &=&-\frac{1}{\hbar }\sum_{n}\int \frac{d^{2}\boldsymbol{k}}{%
\left( 2\pi \right) ^{2}}f_{n}^{0}\left( \boldsymbol{k}\right) \Omega _{n,%
\boldsymbol{k}}=-\sigma _{yx}^{0},  \label{Hall} \\
\alpha _{xy}^{0} &=&\frac{1}{\hbar T}\sum_{n}\int \frac{d^{2}\boldsymbol{k}}{%
\left( 2\pi \right) ^{2}}\Omega _{n,\boldsymbol{k}}\{\frac{\varepsilon
_{n}-\mu }{k_{B}}f_{n}^{0}\left( \boldsymbol{k}\right)  \notag \\
&&-T\log \left[ 1-e^{-\left( \varepsilon _{n}-\mu \right) /k_{B}T}\right] \}
\notag \\
&=&-\alpha _{yx}^{0},  \notag
\end{eqnarray}%
\begin{eqnarray*}
\sigma _{\gamma \gamma }^{1} &=&\frac{\tau }{k_{B}T}\sum_{n}\int \frac{d^{2}%
\boldsymbol{k}}{\left( 2\pi \right) ^{2}}v_{b,\gamma =x,y}^{2} \\
&&\times \left\{ f_{n}^{0}\left( \boldsymbol{k}\right) +\left[
f_{n}^{0}\left( \boldsymbol{k}\right) \right] ^{2}\right\} \\
&=&\sum_{n}\int \frac{d^{2}\boldsymbol{k}}{\left( 2\pi \right) ^{2}}%
f_{\sigma _{\gamma }}, \\
\alpha _{\gamma \gamma }^{1} &=&\frac{\tau }{k_{B}T}\sum_{n}\int \frac{d^{2}%
\boldsymbol{k}}{\left( 2\pi \right) ^{2}}v_{b,\gamma }^{2} \\
&&\times \frac{\varepsilon _{n}-\mu }{k_{B}T}\left\{ f_{n}^{0}\left( 
\boldsymbol{k}\right) +\left[ f_{n}^{0}\left( \boldsymbol{k}\right) \right]
^{2}\right\} , \\
&=&\sum_{n}\int \frac{d^{2}\boldsymbol{k}}{\left( 2\pi \right) ^{2}}%
f_{\alpha _{\gamma }},
\end{eqnarray*}%
where $f_{\sigma _{\gamma }}=\frac{\tau }{k_{B}T}v_{b,\gamma }^{2}\left\{
f_{n}^{0}\left( \boldsymbol{k}\right) +\left[ f_{n}^{0}\left( \boldsymbol{k}%
\right) \right] ^{2}\right\} $, $f_{\alpha _{\gamma }}=\frac{\tau }{k_{B}T}%
v_{b,\gamma }^{2}\frac{\varepsilon _{n}-\mu }{k_{B}T}\left\{ f_{n}^{0}\left( 
\boldsymbol{k}\right) +\left[ f_{n}^{0}\left( \boldsymbol{k}\right) \right]
^{2}\right\} $.

\section{Forbidden of $\protect\sigma _{xy}^{0}\,\left( \protect\alpha %
_{xy}^{0}\right) $ in monolayer 2D semiconductors}

\subsection{Monolayer TMDs}

For a monolayer TMD that consider the intervalley e-h Coulomb exchange
interaction, as shown in Fig. \ref{fig1}(a), the effective exciton
Hamiltonian based on the basis of valley excitons $\left\{ \left\vert 
\boldsymbol{k}\right\rangle _{K},\left\vert \boldsymbol{k}\right\rangle
_{-K}\right\} $ has the form as \cite{Yu,Yu1,Lou,Li} 
\begin{equation}
H_{\mathrm{TMD}}^{ex}=\frac{\hbar ^{2}k^{2}}{2m_{ex}}+J\frac{k}{\left\vert
K\right\vert }\left( 
\begin{array}{cc}
1 & -e^{-2i\varphi } \\ 
-e^{-2i\varphi } & 1%
\end{array}%
\right) ,  \label{H_T}
\end{equation}%
$\boldsymbol{k}=\left( k\cos \varphi ,k\sin \varphi \right) =\left(
k_{x},k_{y}\right) $ is COM momentum of excitons, and $m_{ex}$ is the
isotropic exciton mass which is close to the free electron mass $m_{fe}$.
Taking parameters in monolayer WSe$_{2}$ as an example, it gives $\left\vert
K\right\vert =4\pi /3a\approx 1.26\,\mathrm{\mathring{A}}^{-1}$, $\varphi
=\arctan \left( k_{y}/k_{x}\right) $. $J\approx 1\,\mathrm{eV}$ due to the
first principle calculation. Here we would like to point out that the
exciton enengy at $k=0$, i.e., the optical band gap of the system, has been
omitted in this model (also in the rest of the paper) for simplicity since
it only brings a total energy shift for the system. The eigenstates of $H_{%
\mathrm{TMD}}$ can be expressed as%
\begin{equation}
\left\vert \boldsymbol{k}\right\rangle _{L}=\frac{-e^{-2i\varphi }\left\vert 
\boldsymbol{k}\right\rangle _{K}+\left\vert \boldsymbol{k}\right\rangle _{-K}%
}{\sqrt{2}},\left\vert \boldsymbol{k}\right\rangle _{T}=\frac{\left\vert 
\boldsymbol{k}\right\rangle _{K}+e^{2i\varphi }\left\vert \boldsymbol{k}%
\right\rangle _{-K}}{\sqrt{2}},
\end{equation}%
with $\left\vert \boldsymbol{k}\right\rangle _{\pm K}\approx \sum_{%
\boldsymbol{q}}F_{l=0}\left( \boldsymbol{q}\right) e^{i\boldsymbol{q}\cdot 
\boldsymbol{r}}\left\vert u_{e,\pm K+\boldsymbol{q}+\frac{\boldsymbol{k}}{2}%
}\right\rangle \left\vert u_{h,\mp K-\boldsymbol{q}+\frac{\boldsymbol{k}}{2}%
}\right\rangle $ is the state of $\pm K$ valley excitons. The isotropic
feature of TMDs gives $m_{e/h,x}=m_{e/h,y}=m_{e/h}$ and $m_{e}\approx
m_{h}\approx m_{fe}/2$ as a general situation \cite{Yu,Li}.

After some tedious calculation, one can easily find that $\Omega _{L}=\frac{%
\Omega _{K}^{ex}+\Omega _{-K}^{ex}}{2}=\Omega _{T}$ is zero with $\Omega
_{K}^{ex}=-\Omega _{-K}^{ex}\neq 0$ (see details in \cite{Supp}). This is
consistent with the analysis in Table \ref{Table I}, also the earlier
experimental result that the main contribution of the exciton Hall effect in
monolayer TMDs comes from the trion instead of excitons \cite{Iwa}. On the
other hand, if there is a population difference between excitons in two
valleys (which can be modulated from optoelectronic methods such as optical
pumping to stimulate excitons in the selected valley \cite{Yu,Lou}), the
anomalous exciton Hall effect (seems like the valley Hall effect) may happen
in this system. This has been proved in our recent work related with
non-Hermitian theory of valley excitons in monolayer TMDs \cite{Li1}.

\subsection{Monolayer BP}

For a monolayer BP, only the $\Gamma $ point contributes the valence and
conduction bands as shown in Fig. \ref{fig1}(b), where the exciton
Hamiltonian has a simple form as%
\begin{equation}
H_{\mathrm{BP}}^{ex}=\frac{\hbar ^{2}k_{x}^{2}}{2m_{ex,x}}+\frac{\hbar
^{2}k_{y}^{2}}{2m_{ex,y}}.  \label{H_B}
\end{equation}%
The highly anisotropic feature of this material gives $m_{e,x}\approx
0.17m_{fe}$, $m_{e,y}\approx 1.12m_{fe}$, $m_{h.x}\approx 0.15m_{fe}$, $%
m_{h,y}\approx 6.35m_{fe}$ \cite{Ji}, resulting in the anisotropic exciton
mass $m_{ex,x\left( y\right) }$ \cite{Neto,Gom,Ji}, also the anisotropic
Bohr radius $a_{B,x\left( y\right) }$ \cite{LYang,XWang,Li2}.

However, since the effective electric band-edge model near the $\Gamma $
point in the form \cite{Neto,Gom,Ji}%
\begin{eqnarray}
H_{\mathrm{BP}}^{e} &\simeq &\left( 
\begin{array}{cc}
\Delta /2+\eta _{c}Q_{x}^{2}+\nu _{c}Q_{y}^{2} & \gamma _{1}Q_{x} \\ 
\gamma _{1}Q_{x} & -\Delta /2+\eta _{v}Q_{x}^{2}+\nu _{v}Q_{y}^{2}%
\end{array}%
\right) ,  \notag \\
&&  \label{He_B}
\end{eqnarray}%
$\allowbreak $causing $\left[ H_{\mathrm{BP}}^{e}\left( Q\right) \right]
^{\ast }=H_{\mathrm{BP}}^{e}\left( Q\right) =\sigma _{z}H_{\mathrm{BP}%
}^{e}\left( -Q\right) \sigma _{z}$ ($\gamma _{1}$ is a real number), i.e.,
existence of $\mathcal{T}$ and $\mathcal{S}$ symmetries. This gives zero
Berry curvature $\Omega _{c\left( v\right) }\left( \boldsymbol{k}\right) $,
forbidding the non-zero $\sigma _{xy}^{0}\,\left( \alpha _{xy}^{0}\right) $
shown in Eq. (\ref{Hall}).

\begin{figure}[tbp]
\begin{center}
\includegraphics[width=0.48\textwidth]{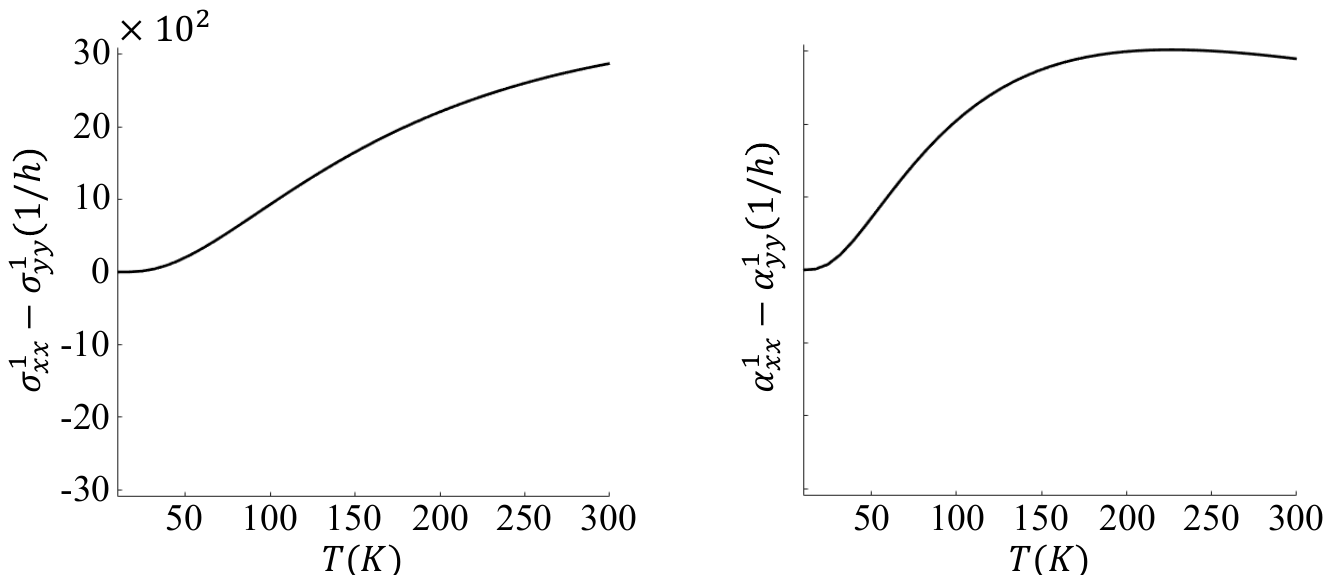}
\end{center}
\caption{(Color online) (a) Numerical calculation of $\protect\sigma %
_{xx}^{1}-\protect\sigma _{yy}^{1}$ in Eq. (\protect\ref{Hall}) for $H_{%
\mathrm{BP}}$ [Eq. (\protect\ref{H_B})], with respect to the the temperature 
$T$. (b) The Similar plot of $\protect\alpha _{xx}^{1}-\protect\alpha %
_{yy}^{1}$ in Eq. (\protect\ref{Hall}) for BP, which has the same y-axis
index as (a). $\protect\mu =-0.01\,\mathrm{eV}$ and $\protect\tau =1\,ps$
for all plots.}
\label{fig2}
\end{figure}

\begin{figure}[tbp]
\begin{center}
\includegraphics[width=0.48\textwidth]{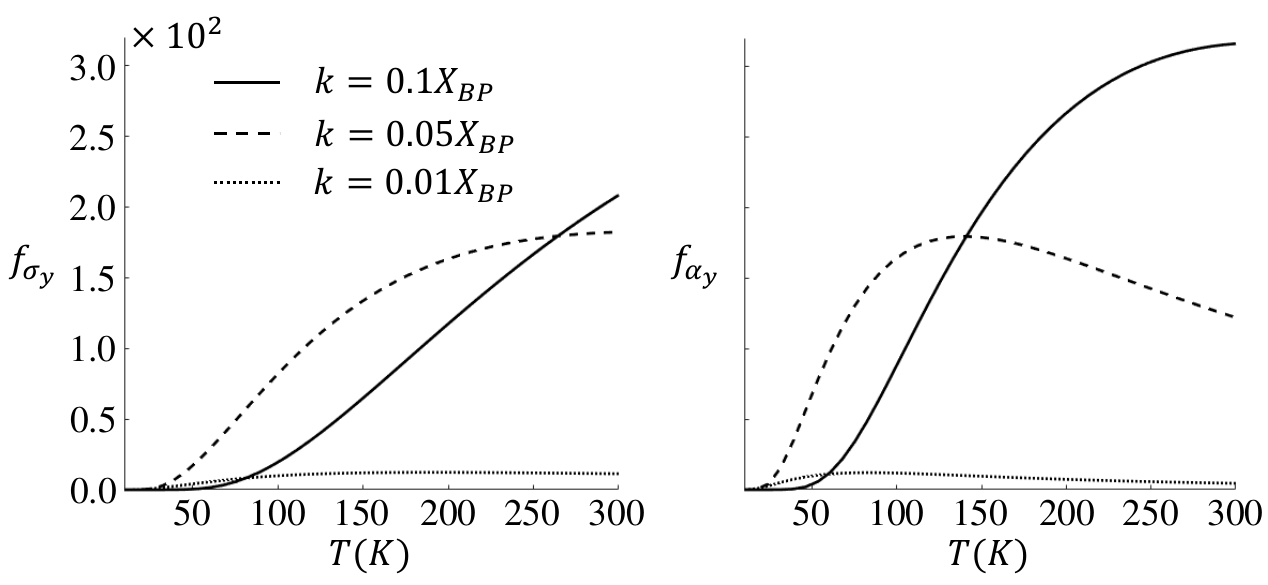}
\end{center}
\caption{(Color online) (a) Numerical calculation of $f_{\protect\sigma %
_{y}} $ in Eq. (\protect\ref{Hall}) for $H_{\mathrm{BP}}$ [Eq. (\protect\ref%
{H_B})] at different values of $\boldsymbol{k}=(k_{x},k_{y})=k(cos\protect%
\varphi ,sin\protect\varphi )$, with respect to the the temperature $T$. (b)
The Similar plot of $f_{\protect\alpha _{y}}$ in Eq. (\protect\ref{Hall})
for BP, which has the same y-axis index as (a). $X_{BP}=\protect\pi /4.58$
means the half width of the first Brillouin zone of BP in $x$ direction. $%
\protect\mu =-0.01\,\mathrm{eV}$ and $\protect\tau =1\,ps$ for all plots.}
\label{fig3}
\end{figure}

\section{Linear exciton Hall (Nernst) effect without Berry curvature in
monolayer 2D semiconductors}

When the exciton Berry curvature is zero in the 2D system, the linear
exciton Hall (Nernst) current is directly connected to the net difference of 
$\sigma _{xx}^{1}-\sigma _{yy}^{1}\,\left( \alpha _{xx}^{1}-\alpha
_{yy}^{1}\right) $ according to the Eq. (\ref{Hall}), which requires
anisotropic exciton masses. In monolayer TMDs, the band-edge model $H_{%
\mathrm{TMD}}^{e}$ typically yields isotropic effective masses of electrons
and holes, i.e., $m_{e,x}=m_{e,y}=m_{e}$, $m_{h,x}=m_{h,y}=m_{h}$, where $%
m_{e}$ can differ slightly from $m_{h}$ \cite{Lou,Lou1,Yak,Giu,Fal,Fal1}. As
a result, $\sigma _{xx}^{1}-\sigma _{yy}^{1}=0\,\left( \alpha
_{xx}^{1}-\alpha _{yy}^{1}=0\right) $ in the effective exciton Hamiltonian (%
\ref{H_T}),\ since the exciton mass is isotropic, which is consistent with
the symmetry analysis from crystal symmetry in Table \ref{Table I}.

In contrast, the effective exciton Hamiltonian $H_{\mathrm{BP}}$ (\ref{H_B})
exhibits different structure due to the strong anisotropy of these
materials, represented by a single parabolic exciton band with a large
disparity between $m_{ex,x}$ and $m_{ex,y}$. A straightforward calculation
reveals that BP has a significant value of $\sigma _{xx}^{1}-\sigma
_{yy}^{1}\,\left( \alpha _{xx}^{1}-\alpha _{yy}^{1}\right) $, corresponding
to a non-zero linear exciton Hall (Nernst) effect even without Berry
curvature, as shown in Fig. \ref{fig2}. Then we demonstrate that the
temperature-dependent behavior of this Hall (Nernst) effect in Fig. \ref%
{fig2} is governed by the term $f_{n}^{0}\left( \boldsymbol{k}\right) +\left[
f_{n}^{0}\left( \boldsymbol{k}\right) \right] ^{2}$ within the integral in
Eq. (\ref{Hall}), as illustrated in Fig. \ref{fig3}. From the definition in
Eq. (\ref{Hall}), we find that $f_{\sigma _{x}}/f_{\sigma _{y}}\,\left(
f_{\alpha _{x}}/f_{\alpha _{y}}\right) =\left( m_{ex,y}/m_{ex,x}\right) ^{2}$%
, which is approximately $545$ in BP. This indicates that $\sigma
_{xx}^{1}\,\left( \alpha _{xx}^{1}\right) $ predominantly contributes to the
linear exciton Hall (Nernst) effect in BP, which are consistent with the
plots in Fig. \ref{fig2}.

In the context of the exciton Hall or Nernst effect, the in-plane driving
force $\boldsymbol{\mathcal{F}}=\hbar \frac{d\boldsymbol{k}}{dt}$ can arise
from either statistical sources, such as gradients in chemical potential ($%
\nabla \mu $) and particle density, or mechanical forces, such as gradients
in particle energy. For example, in the experiment \cite{Iwa}, laser
illumination on a sample edge was used to create temperature ($\nabla T$)
and/or the chemical potential ($\nabla \mu $) gradients, driving the
diffusive transport of excitonic quasiparticles. To generate a mechanical
force for neutral excitons, one could consider introducing a strain
gradient, as strain modifies the bandgap and thus affects exciton energy.
Consequently, the effect we showed here could be explored at finite
temperatures by applying an appropriate in-plane driving force along
specific directions in anisotropic monolayer 2D semiconductors.

\section{Dissussion and conclusions}

Recent studies indicate that, under certain circumstance, such as the
sufficiently large temperature and relatively strong scatters,
skew-scattering and especially side-jump processes due to the exciton
scattering by an impurity or phonon can significantly affect the exciton
valley Hall effect, where the side-jump effect can effectively compensate
the Berry-curvature-induced anomalous velocity \cite{Gol,Gol1,Dya}. The
detailed calculation shows the valley Hall current of excitons are still in
linear form of $\boldsymbol{\mathcal{F}}$, $\nabla T$ and $\nabla \mu $, but
the conductivity is dominated by the contribution of skew scattering and
approximately proportional to $T$ \cite{Gol}. By contrast, when the
temperature is low and the scattering process is well suppressed,
Berry-curvature-induced anomalous velocity can remain operative even in the
presence of side jump, which is consistent with the established theory and
experiments on the anomalous/valley Hall effects in electronic systems \cite%
{Niu6,Niu7,Niu,McE,Tak,Tar,AGao}. This implies that\ our low-temperature
analysis of the exciton Hall effect--and our treatment of the exciton Nernst
effect focusing on Berry curvature--remain applicable.

In summary, we have demonstrated that the linear exciton Hall (Nernst)
effect exists in monolayer 2D semiconductors. The strong anisotropy of
certain 2D materials can result in significant exciton Hall (Nernst)
conductivity, even in the absence of exciton Berry curvature. This effect
leads to enhanced excitonic emission at the sample boundaries, presenting an
exciting opportunity for future experimental investigations. In the presence
of disorder, the excitonic Hall effect can still be visualized as a curved
transport trajectory of exciton emission in polarization-resolved
photoluminescence mapping, offering a practical route for future
experimental investigations. Conversely, in the recently studied 2D
materials we considered, the Hall (Nernst) effect driven by non-zero exciton
Berry curvature is strictly forbidden by $\mathcal{T}$ and spatial
symmetries. Thus, the question of whether stable and easily fabricable
monolayer 2D materials can support a net exciton Hall (Nernst) effect with
non-zero exciton Berry curvature remains open. Addressing this question is
an intriguing direction for future research.

\textit{Acknowledgments.}---C. Li would like to thank Dawei Zhai, Bo Fu, and
B. Bao for useful discussions, helpful suggestions from Wang Yao, and the
support from the National Natural Science Foundation of China (12504205),
the Fundamental Research Funds for Young Scholars of Jiangsu Province
(BK20250344), and the Fundamental Research Funds for the Central
Universities from China. Q. Tong would like to thank the support from the
National Key Research and Development Program of Ministry of Science and
Technology (2022YFA1204700, 2021YFA1200503), the National Natural Science
Foundation of China (12374178), the Science Fund for Distinguished Young
Scholars of Hunan Province (2022J10002), and the Fundamental Research Funds
for the Central Universities from China.


\end{document}